\documentclass[12pt]{iopart}

\usepackage{iopams}
\usepackage{bbm}

\usepackage{graphicx} 
\usepackage{subfigure}

\newcommand{\s}[1]{\ensuremath{\sum_{#1}}}

\newcommand{\Jij}{{\ensuremath{J_{ij}}}}
\newcommand{\Jijp}[1]{{\ensuremath{J^{#1}_{ij}}}}

\newcommand{\nui}{\ensuremath{\nu_i}}

\newcommand{\qsi}{\ensuremath{\tilde{q}_i}}
\newcommand{\qsj}{\ensuremath{\tilde{q}_j}}
\newcommand{\qmi}{\qsi - m_i^2}
\newcommand{\qmj}{\qsj - m_j^2}

\newcommand{\qs}{\ensuremath{\tilde{q}}}

\newcommand{\mi}[1]{\ensuremath{\langle#1\rangle{}}}

\newcommand{\xLnx}[1]{\ensuremath{({#1})\ln\left({#1}\right)}}
\newcommand{\xLnxh}[1]{\xLnx{\frac{{#1}}2}}

\newcommand{\ijp}{{(ij)}}
\newcommand{\ij}{{ij}}

\newcommand{\Hop}{\mathcal{H}}
\newcommand{\Frei}{\mathcal{F}}
\newcommand{\Gibbs}{\mathcal{G}}

\newcommand{\T}{\mathbb{T}}

\newcommand{\strike}[1]{\ensuremath{\text{\textit{#1}\kern-0.8em\hbox{$\diagdown$}}}}

\newcommand{\vx}{\vec{x}}

\newcommand{\miI}[1]{\mi{#1}_I}
\newcommand{\qsh}{\kappa}%
\newcommand{\udl}[1]{\underline{{#1}}}
\newenvironment{bmatrix}{\left(\begin{array}{ll}}{\end{array}\right)}

\begin{document}

\jl{3}
\title{Fermionic TAP-equations}
\author{Martin Rehker and Reinhold Oppermann}
\address{Institut f. Theoret. Physik, Univ. W{\"u}rzburg, D-97074 W{\"u}rzburg, Federal Republic of Germany}

\begin{abstract}
  We derive the TAP equations for the fermionic Ising spin glass.  It
  is found that, just as in the non-fermionic model, the conditions
  for stability and for validity of the free energy are equivalent.
  We determine the breakdown of the paramagnetic phase. Numeric
  solutions of the fermionic TAP-equations at $T=0$ allowed to
  localize a first order transition between the spin glass phase and
  the paramagnetic phase at $\mu \approx 0.8$. We computed at
  zero temperature the filling factor $\nu(\mu)$ and the distribution
  of internal fields. The saddle-point equations resulting from the
  calculation of the number of solutions to the TAP-equations were
  found to be much more complicated as in the non-fermionic case.
\end{abstract}

\submitted

\maketitle

\section*{Introduction}
\label{introduction}

Thouless, Anderson and Palmer derived local mean-field
equations for the random Ising model \cite{tap-76}.  We generalize
these TAP-equations to the four-state fermionic spin glass including a
chemical potential $\mu$. $\tilde{n}$ is the number operator, $\Jij$ are
gaussian distributed random interactions with variance $J^2/N$:

\begin{equation*}
    \Hop = -\frac 1 2 \sum_\ij\Jij \sigma_i \sigma_j - \mu \sum_i \tilde{n}_i - \sum_i h_i^{ext} \sigma_i.
\end{equation*}

It is believed \cite{dasgupta-sompolinsky-83,sompolinsky-81}, that the
complete set of solutions to the TAP-equations is equivalent to the
fully replica symmetry broken solution in the quantum-field theory
\cite{oppermann-96,oppermann-a-98,oppermann-b-98,oppermann-92}, so far
only known for half filling. We were able to solve the fermionic
TAP-equations numerically on the $T=0$ axis and to determine the
first order transition at $\mu_c \approx 0.8$ where the paramagnetic-
and spin glass free energies become equal. The dependence of the
filling factor on the chemical potential and the distribution of
internal fields are given below.

\section{Fermionic TAP-equations}
\label{sec:fermionic-tap-equations}

The free energy corresponding to the non-fermionic TAP-equations is:

\begin{eqnarray}
  \eqalign{
  \label{eq:non-fermionic-free-energy}
  \Frei =~& -\sum_i h_i^{ext} m_i - \frac 1 2 \sum_\ij \Jij m_i m_j - 
  \frac{\beta}4 \sum_\ij \Jijp{2} (1 - m_i^2)(1 - m_j^2) \\ &+
  \frac 1 \beta \s{i} \Bigl\{ 
    (1 + {m_i}) \ln(\frac{1 + {m_i}} 2) + 
    (1 - {m_i}) \ln(\frac{1 - {m_i}} 2) \Bigr\} }
\end{eqnarray}

In order to calculate the corresponding expression for the
generalized fermionic model, we extended the
linked cluster diagrammatic theory by Horwitz and Callen
\cite{horwitz-callen-61}. For the sake of simplicity this rigorous and
lengthy derivation is replaced by shorter and more intuitive
arguments.

The terms involving logarithms in equation
(\ref{eq:non-fermionic-free-energy}) correspond to the entropy of an
ensemble of spins in the non-fermionic two state model with relative
occupations $n_{i\uparrow}$ and $n_{i\downarrow}$, using $m_i
=n_{i\uparrow} - n_{i\downarrow}$. Now we replace these terms with the
entropy of the ensemble in the extended fermionic four state model.
The relative occupations are denoted by $n_{i\uparrow}$,
$n_{i\downarrow}$, $n_{i0}$ and $n_{i\uparrow\downarrow}$, setting
$m_i =n_{i\uparrow} - n_{i\downarrow}$, $n_i = n_{i\uparrow} +
n_{i\downarrow} + 2\, n_{i\uparrow\downarrow}$ and $\qsi =
n_{i\uparrow} + n_{i\downarrow}$.  Then we account for the non-trivial
occupation number of the magnetic states by replacing $1 - m_i^2$ in
the Onsager reaction field by $\qmi$. After a final Legendre
transformation the fermionic free energy reads

\begin{eqnarray}
  \fl\eqalign{
  \Frei =~& - \frac 1 2 \sum_\ij \Jij m_i m_j - \frac {\beta}4 \sum_\ij \Jij
  (\qmi)(\qmj) - \sum_i \mu (1 + (1 - \qsi)\,\T)
    + \frac{N}{\beta} \ln 2 \\ &+ 
    \frac 1{\beta} \sum_i \Bigl[
    \xLnxh{\qsi+m_i} + \xLnxh{\qsi-m_i}  \\ &+
    \xLnxh{(1-\qsi)(1-\T)} \\ &+ \xLnxh{(1-\qsi)(1+\T)} 
    \Bigl]  
  \label{eq:freie-energie}}
\end{eqnarray}

The system is characterized by the following $2 \, N$ coupled TAP-equations.

\begin{equation}
  \eqalign{
  \label{tap-isgf}
  {m_i} &= \frac{\sinh(\beta H_i)}{\cosh(\beta H_i) + 
    \cosh(\beta \mu) \exp(-\beta X_i)} \\
  \qsi &= \frac{\cosh(\beta H_i)}{\cosh(\beta H_i) + 
    \cosh(\beta \mu) \exp(-\beta X_i)}},
\end{equation}

where $H_i = h_i^{ext} + \s{j} \Jij m_j -
\beta {m_i} \s{j} \Jijp{2} (\qmj)$ and $X_i = \frac{\beta}{2} \s{j}
\Jijp{2} (\qmj)$. A third equation for the local filling factors $\nui
= 1 + (1-\qs_i)\tanh(\beta\mu)$ follows from $\nu = - \partial_\mu
\Gibbs$. In the replicated quantum-field theory the corresponding equation $\nu =
1 + (1 - \qs) \tanh(\beta \mu)$ turns out to be invariant under an
arbitrary number of replica symmetry breaking steps.

\section{Convergence and Stability}
\label{sec:konvergence-and-stability}

Inspired by Plefka's work on the non-fermionic system \cite{plefka-82}
and adopting his replacement $\Jij \rightarrow \alpha \Jij$ we can
rederive the fermionic free energy by a Taylor expansion as follows

\begin{eqnarray}
  \fl\eqalign{
  \label{tapfealpha}
  \Gibbs(\alpha) =~& -\frac{\alpha}{2} \sum_\ij \Jij m_i m_j - 
  \alpha^2 \frac \beta 4 \sum_\ij \Jijp{2} (\qmi)(\qmj)
  \\ &+
    \frac 1{\beta} \sum_i \Bigl[
    \xLnxh{\qsi+m_i} + \xLnxh{\qsi-m_i}\\ &+
    \xLnxh{(1-\qsi)(1+\T)} \\ &+ 
    \xLnxh{(1-\qsi)(1-\T)} 
    \Bigl]  \\ &- 
    \mu \sum_i (1 + (1-\qsi)\,\T)
    - \sum_i h_i^{ext} m_i + \frac{N}{\beta} \ln 2 + O(\alpha^3) .}
\end{eqnarray}

The correct free energy is given by $\Gibbs(\alpha=1)$. The diagrammatic
expansion showed already that, provided the series actually converges,
the terms of cubic order or higher are suppressed in the
thermodynamic limit.  We may now determine the radius of convergence
of $\Gibbs(\alpha)$, which, by a standard theorem of complex analysis,
is equivalent to the radius of convergence of

\begin{equation*}
  \partial_\alpha \Gibbs(\alpha) =  -\frac{1}2 \mi{\sum_\ij\Jij \sigma_i \sigma_j}_{\alpha} = -\frac1 2 \sum_\ij \Jij m_i m_j -
  \frac 1 {2\beta} \sum_\ij \Jij \chi_{ij}(\alpha).
\end{equation*}

There, the susceptibility matrix $\chi$ is defined as

\begin{equation*}
  \chi_{ij}(\alpha) = \beta( \mi{\sigma_i \sigma_j}_{\alpha} - m_i m_j )
  = \partial_{h_i} \partial_{h_j} \Frei(\udl{h},\udl{\chi}).
\end{equation*}

The radius of convergence is given by $\min(|\alpha|)$, where the
minimization is done over the values of $\alpha$ with eigenvalues $0$
in the inverse susceptibility matrix $\chi^{-1}$. In the
non-fermionic case the equation $(\chi^{-1})_{ij} = (\left[ \partial_{\udl{h}}
  \partial_{\udl{h}} \Frei(\udl{h}) \right]^{-1})_{ij} = \partial_{m_i}
\partial_{m_j} \Gibbs(\udl{m})$ signifies that, when taking into account the
special properties of the spectra of these random matrices
\cite{mehta}, the local stability of a TAP-solution implies the
validity of the free energy at this point. In the fermionic model we
have to deal with two different matrices to describe either the
convergence of the linked cluster expansion $(\chi^{-1})_{ij}$ or the local
stability of a given TAP-solution by

\begin{equation*}
  \begin{bmatrix}
    \partial_{m_i} \partial_{m_j} \Gibbs &     \partial_{m_i} \partial_{\qsj} \Gibbs\\
    \partial_{\qsi} \partial_{m_j} \Gibbs &     \partial_{\qsi} \partial_{\qsj} \Gibbs 
  \end{bmatrix}  = 
  \begin{bmatrix}
    \partial_{h_i} \partial_{h_j} \Frei &     \partial_{h_i} \partial_{\chi_j} \Frei \\
    \partial_{\chi_i} \partial_{h_j} \Frei &     \partial_{\chi_i} \partial_{\chi_j} \Frei
  \end{bmatrix}^{-1}.
\end{equation*}

A theorem by Pastur \cite{pastur-72,pastur-74} used heavily by Plefka
\cite{plefka-82} and resolvent-calculus can be applied to both cases
to determine the limits of the support of the spectra. It is very
interesting to note, that at the end of quite lengthy calculations
both matrices lead exactly to the same set of conditions generalizing
Plefka's convergence and stability conditions by

\begin{eqnarray}
  \mi{(\qs - m^2)^2} & \leq T^2 \label{eq:stability-one}\\
  \frac 1 2 \mi{\qs(1-\qs)} + 2 \mi{m^2 - m^4} & \leq T^2 .
  \label{eq:stability-two}
\end{eqnarray}

The known non-fermionic limit is obtained by setting $\qsi = 1$.  We
have thus linked the local stability of TAP-solutions to the finite
support of the spectrum of random matrices in the thermodynamic limit.
This gives a hint why the numerical search for these solutions is so
difficult
\cite{ling-bowman-levin-83,soukoulis-grest-levin-a-83,soukoulis-grest-levin-b-83}:
for finite $N$ the support of these spectra becomes unbounded, see
e.g. \cite{mehta} for the exponential corrections to the
semi-circle-law at finite $N$. This means, that for every solution the
probability of having \emph{negative} eigenvalues in the
stability matrix is finite.  But already one negative eigenvalue
prevents finding this solution via minimization algorithms. Most
solutions become thus unstable.

\section{Breakdown of the homogeneous paramagnetic solution}
\label{sec:homogeneous-paramagnetic-solution}

The TAP-equations are readily solved numerically in the homogeneous
paramagnetic phase, where they reduce to one single equation for $\qs
= \qsi$ given for all $i$ by

\begin{equation}
  \label{eq:homogene-selbstkonsistenzgleichung}
  \qs = \frac 1 {1 + \cosh(\beta\mu) \exp(-\frac{(\beta J)^2}2 \qs)} .
\end{equation}

This is exactly the equation found in
\cite{oppermann-96,oppermann-92,sachdev-read-oppermann-95} for $\qs$
with a replicated quantum-field theory approach.

\begin{figure}[hp]
  \begin{center}
    \includegraphics{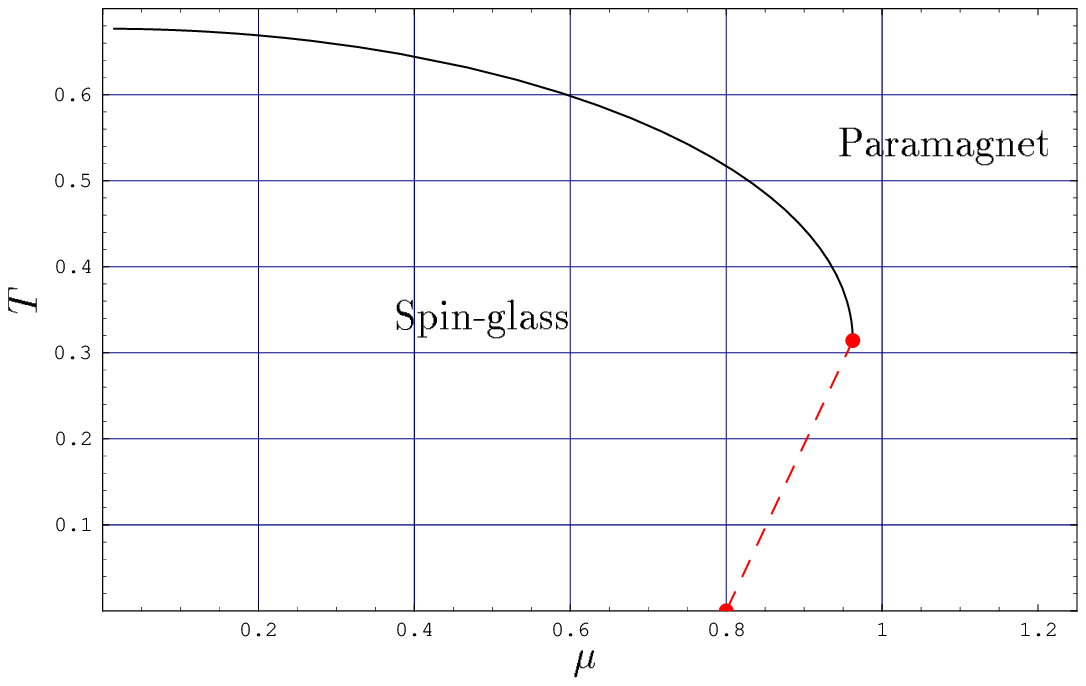}
    \caption{Phase diagram obtained from the fermionic TAP-equations. The dotted line is 
      a linear approximation th the first order transition line connecting the (calculated)
      tricritical point and the $T=0$ critical point.}
    \label{fig:phasediagram}
  \end{center}
\end{figure}

The second-order transitions between the paramagnetic and the
spin glass phase are given by the intersection of the paramagnetic
solutions with the stability conditions (equations
(\ref{eq:stability-one}) and (\ref{eq:stability-two})). In the next
section we find a first order transition on the $T=0$ axis at $\mu_c \approx
0.8$. We expect a line of first order transitions linking this point
with the tricritical point. The dotted line in the
phase diagram (figure \ref{fig:phasediagram}) gives only a schematic
behaviour of this line as the exact path is yet unknown.

\section{TAP-equations at $T=0$}
\label{sec:tap-equations-at-t-zero}

\begin{figure}[hp]
  \begin{center}
   \includegraphics{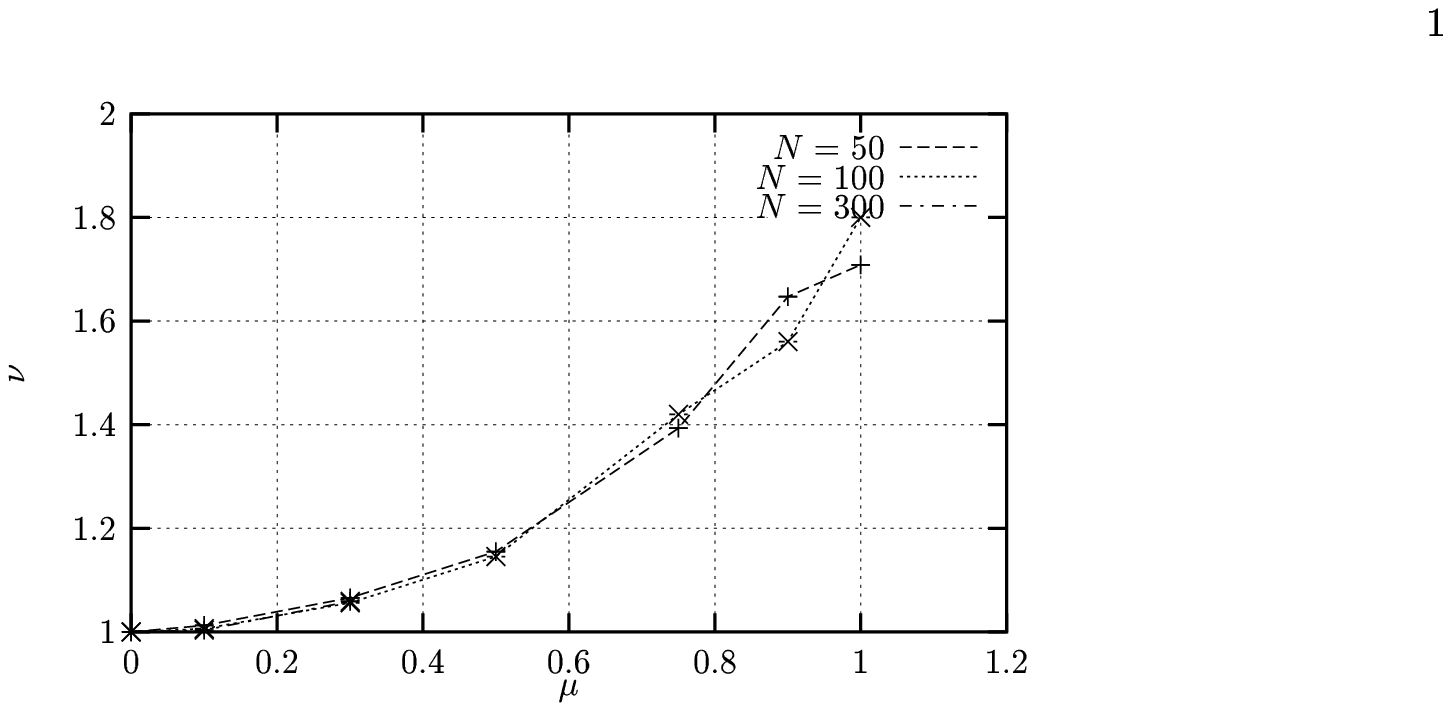}   
    \caption{Filling factor $\nu(\mu)$ as a function of the 
      chemical potential for systems of different size.}
    \label{fig:global}
  \end{center}
\end{figure}

At $T = 0$ the TAP-equations reduce to 

\begin{equation*}
  m_i = \Omega( h_i ), \quad  \qsi = m_i^2 
\end{equation*}

where $h_i = \sum_i \Jij m_i$ and $\Omega(x) = \Theta(x - \mu) -
\Theta(\mu - x)$ denotes a modified sign function. The energy
corresponding to these solutions is simply

\begin{equation*}
  f_{SG} = - \sum_\ijp \Jij m_i m_j - \mu \sum_i n_i
\end{equation*}

which has to be compared with the free energy of the paramagnetic
solution $f_{PM} = - 2 \mu$ to find the first order transition. We were
able to calculate numerically a huge number of spin glass solutions.
We first note the interesting dependence of the filling factor $\nu$
on the chemical potential (see figure \ref{fig:global}). Unlike 
the discontinuous replica-symmetric and finite-step RSB solutions
\cite{oppermann-a-98,oppermann-b-98} the filling factor 
varies continously with $\mu$ in the vincinity of $\mu = 0$.
The numerical data for the increase of $\nu(\mu)$ near $\mu = 0$ are
compatible with power law fits $|\delta\nu| \propto |\mu|^x$, $x > 1$,
including exponential behavior.

\begin{figure}[hp]
  \begin{center}
    \subfigure[$N=75$]{\includegraphics{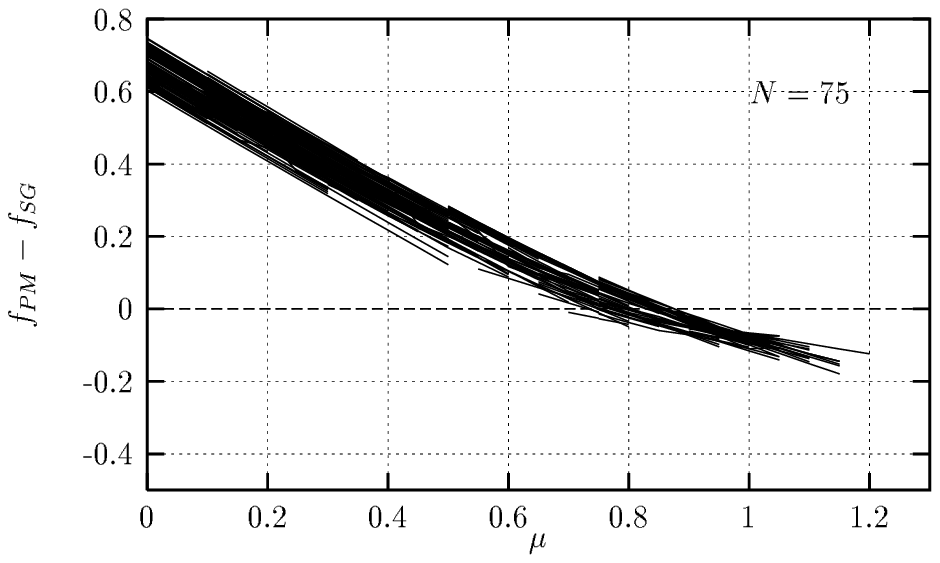}}
    \subfigure[$N=150$]{\includegraphics{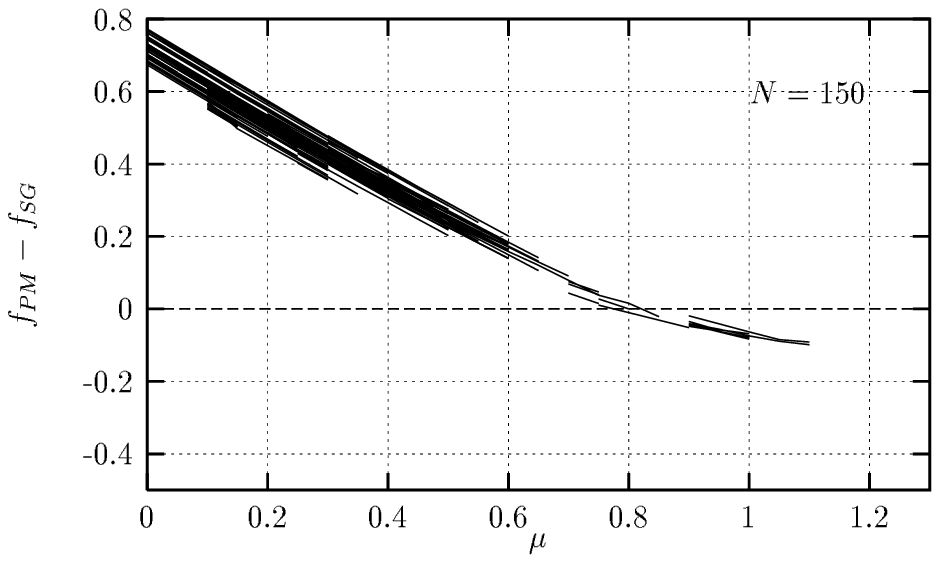}}
    \caption{Energy gap between spin glass solutions of the zero temperature 
      TAP-equations and the paramagnetic solution for systems of different
      size.}
    \label{fig:energy-gap}
  \end{center}
\end{figure}

From the dependence of the energy difference between the spin glass
solutions and the paramagnetic solution (figure \ref{fig:energy-gap})
we can deduce a first order transition at $\mu = \mu_{c1} \approx 0.8$.
This critical value of $\mu$ can be viewed as a $T=0$ analogue of the $T_{c1}$
for the thermal first order transitions.

\begin{figure}[hp]
  \begin{center}
    \resizebox{11cm}{!}{\includegraphics{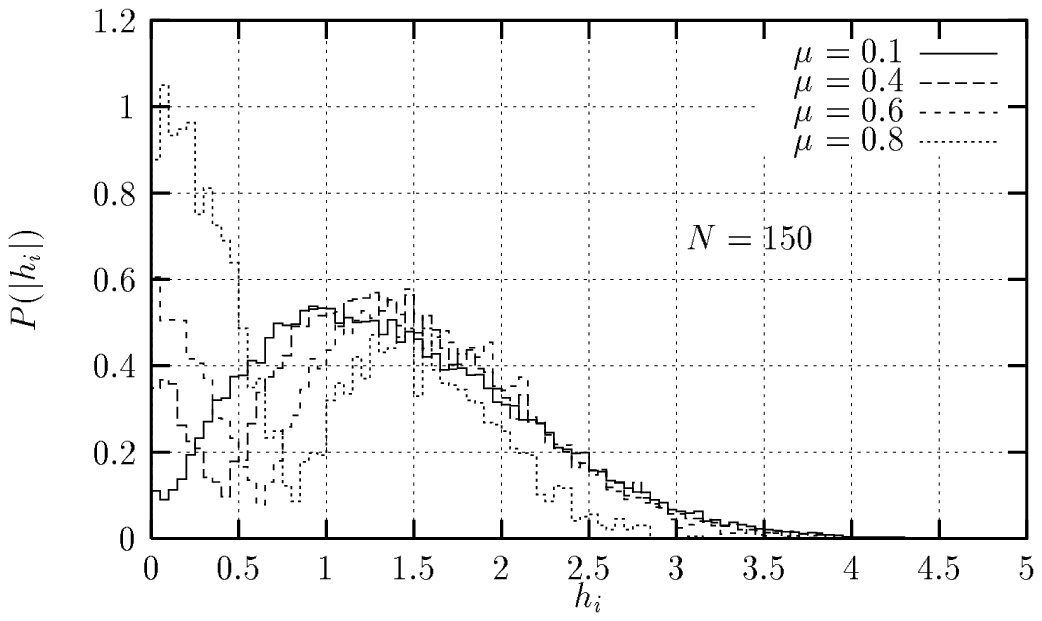}}      
    \caption{Internal field distribution $P(|h_i|)$ 
      for $N=150$ and different values of $\mu$. The distribution
      $P(h_i)$ is symmetric. We used about $10000$ points for each
      histogram.}
    \label{fig:h-distribution}
  \end{center}
\end{figure}

\begin{figure}[hp]
  \begin{center}
    \resizebox{11cm}{!}{\includegraphics{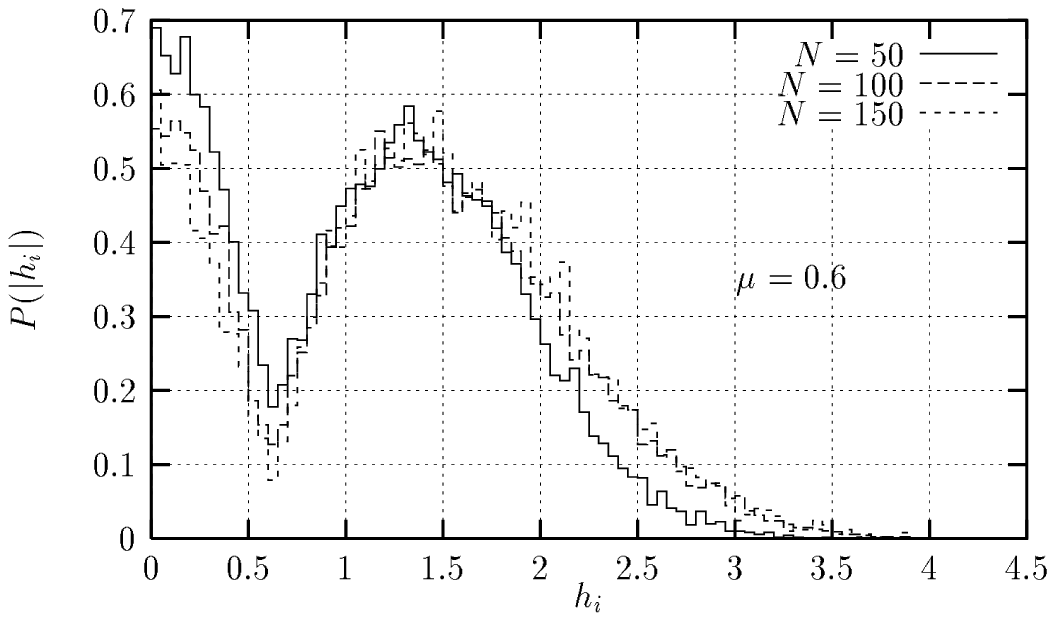}}      
    \caption{Scaling of $P(h_i)$ at $N=50$, $N=100$, and
      $N=150$ for $\mu = 0.6$.}
    \label{fig:distribution-h-int-scale}
  \end{center}
\end{figure}

Another very interesting feature shows up in the behaviour of the
distribution of the local fields $h_i$. When applying a chemical
potential this probability density function is substantially modified.
The ``softgap'' (see \cite{roberts-81}) at $h_i=0$ splits up into two softgaps
at $h_i = -\mu$ and $h_i = \mu$. Within the interval $[-\mu,\mu]$
another peak emerges (see figures \ref{fig:h-distribution} and
\ref{fig:distribution-h-int-scale}).

\section{Number of solutions}
\label{sec:number-of-solutions}

The number of solutions of the TAP-equations (related to the so-called
complexity) can be calculated adopting the procedure of Bray and Moore
\cite{bray-moore-80} for finite $T$ or following Roberts \cite{roberts-81} for
$T=0$. We were able to obtain in both cases fermionic generalizations of the
saddle-point equations, which become extremely complicated due to the
additional non-magnetic degrees of freedom. For example in the
finite temperature case the equations for the parameters $q$, $\qs$
$\eta$, $\varrho$, $\Delta$ and $B$ read

\begin{equation}
  \label{eq:msr-funktional-sattelpunktsgleichungen}
  \eqalign{
    f &= \miI{\Frei}, \qquad
    q = \miI{m^2}, \qquad
    \qs = \miI{\qsh} \\
    0 &= B \left\{ 1 - 
      J^2 \miI{\frac{(\qsh - (\qsh\lambda)^2)^2}{1 + B(\qsh - (\qsh\lambda)^2)}}
    \right\} \\
    0 &= \frac 1 {q} 
    \miI{\qsh\lambda(\tanh^{-1}(\lambda) - \qsh\lambda\Delta)}
    -  \Delta  - J^2 (\qs - q) \\
    \varrho &= 
    -(\Delta + B) -\frac{J^2}2 \miI{\Theta\sqrt{1-\lambda^2}\qsh^2\Xi} \\
    \eta &= \Delta + B + 
    \frac 1 {2 q} \left[
      1 -
      \frac 1 {J^2 q} \miI{(tanh^{-1}(\lambda) - \qsh\lambda\Delta)^2}
    \right] 
     + \frac{J^2}2 \miI{\Theta\sqrt{1-\lambda^2}\qsh^2\Xi} ,}
\end{equation}

which can be viewed as an extension of the original equations
given by Bray and Moore, but the average $\miI{\cdot}$ means averaging
by use of the following kernel:

\begin{eqnarray*}
    I =~&\int_{0}^1 d\qsh \int_{-1}^{1} d \lambda \,
    g(\vx,\qsh,\lambda) \delta( f(\vx,\qsh,\lambda) ) \\ 
    =&\int_{0}^1 d\qsh \int_{-1}^{1} d \lambda \,
    \frac {\qsh}{\sqrt{2\pi} \sqrt{q} J} \,
    \left[ \frac {1 + B (\qsh - (\qsh \lambda)^2)}%
      {(\qsh^2 - (\qsh \lambda)^2)} \right] (1-\qsh)^{u-1} \\& \times
    \exp\Bigl(\eta\,(\qsh \lambda)^2 +
      \varrho\,\qsh + \frac{u\,\qsh\,\lambda}2 \tanh^{-1}(\lambda)  -
      \frac{1}{2 J^2 q}\,\left( \tanh^{-1}(\lambda) - \qsh \,\lambda\,\Delta
      \right)^2 \Bigr)
    \\ &
    \times \delta\left(
      \ln\left(\frac{(\qsh^2 - (\qsh \lambda)^2)\cosh^2(\mu)}{(1-\qsh)^2}\right)
      - J^2 (\qs - q)
    \right)
\end{eqnarray*}

$\lambda = m/\qsh$ is the reduced magnetisation, $\Xi$ an abbrevation for a
very lengthy expression. These equations remain currently unsolved even numerically.
The equations for $T=0$ are equally
hard to treat.

\section{Outlook}
\label{sec:conclusions}

The numerical work presented here should now be accompanied by exact
analytic RSB calculations for arbitrary $\mu$. In order to reproduce
the behavior of $\nu(\mu)$ and of $P(h_i,\mu)$ the generalizations of
Robert's saddle-point equations \cite{roberts-81} should be solved. It
would be desirable to have Parisi's solution (infinite step RSB) for
$\mu \neq 0$. The answer to one of the open questions might reveal the
exact path of the first order transition line for $T\neq0$.  If this
turns out to be impossible, one should find more refined numeric
algorithms, which allow to solve the fermionic TAP-equations for
$T\neq 0$.


\newpage

\begin{thebibliography}{10}

\bibitem{bray-moore-80}
A.~J. Bray and M.~A. Moore.
\newblock {\em J. Phys. C}, 13:L469--76, 1980.

\bibitem{dasgupta-sompolinsky-83}
C.~Dasgupta and H.~Sompolinsky.
\newblock {\em Phys. Rev. B}, 27(7):4511--4514, 1983.

\bibitem{horwitz-callen-61}
Gerald Horwitz and Herbert~B. Callen.
\newblock {\em Phys. Rev.}, 124(6):1757--1785, 1961.

\bibitem{ling-bowman-levin-83}
David~D. Ling, David~R. Bowman, and K.~Levin.
\newblock {\em Phys. Rev. B}, 28(1):262--269, 1983.

\bibitem{mehta}
Mandan~Lal Mehta.
\newblock {\em Random Matrices}.
\newblock Academic Press, 2 edition, 1991.

\bibitem{oppermann-96}
R.~Oppermann and B.~Rosenow.
\newblock In {\em Complex Behavior of Glassy Systems}, Sitges, 5 1996.
\newblock cond-mat/9610055.

\bibitem{oppermann-a-98}
R.~Oppermann and B.~Rosenow.
\newblock {\em submitted to Phys. Rev. B}, 1998.
\newblock cond-mat/9803239.

\bibitem{oppermann-b-98}
R.~Oppermann and B.~Rosenow.
\newblock {\em submitted to Phys. Rev. B}, 1998.
\newblock cond-mat/9803249.

\bibitem{oppermann-92}
Reinhold Oppermann and Axel M{\"u}ller-Groeling.
\newblock {\em Nuclear Physics B}, 401:507--547, 1993.

\bibitem{pastur-72}
L.~A. Pastur.
\newblock {\em Theoret. Mat. Fiz.}, 10:102--112, 1972.

\bibitem{pastur-74}
L.~A. Pastur.
\newblock {\em Russ. Math. Surveys}, 28:1--67, 1974.

\bibitem{plefka-82}
T.~Plefka.
\newblock {\em J. Phys. A}, 15:1971--1978, 1982.

\bibitem{roberts-81}
Stephen~A. Roberts.
\newblock {\em J. Phys. C}, 14:3015, 1982.

\bibitem{sachdev-read-oppermann-95}
Subir Sachdev, N.~Read, and R.~Oppermann.
\newblock {\em Phys. Rev. B}, 52(14):10286--94, 1995.

\bibitem{sompolinsky-81}
H.~Sompolinsky.
\newblock {\em Phys. Rev. Let.}, 47:935, 1981.

\bibitem{soukoulis-grest-levin-a-83}
C.~M. Soukoulis, G.~S. Grest, and K.~Levin.
\newblock {\em Phys. Rev. B}, 28(3):1495--1509, 1983.

\bibitem{soukoulis-grest-levin-b-83}
C.~M. Soukoulis, G.~S. Grest, and K.~Levin.
\newblock {\em Phys. Rev. B}, 28(3):1510--1523, 1983.

\bibitem{tap-76}
D.~J. Thouless, P.~W. Anderson, and R.~G. Palmers.
\newblock {\em Philosophical Magazine}, 35(3):593--601, 1977.

\end{thebibliography}
\end{document}